\documentclass[submit,5p]{elsarticle}

\usepackage{amssymb}
\usepackage{amsmath}
\usepackage{lineno}
\usepackage{xcolor}
\usepackage{float}
\usepackage{pifont} % special characters
\usepackage{siunitx} % should work: https://tex.stackexchange.com/questions/124212/siunitx-and-elsevier-ees
\usepackage{url}

\graphicspath{{figures/}}
\journal{Elsevier}

\begin{document}
	
	\begin{frontmatter}
		
		\title{Improving the Spatial Resolution of Silicon Pixel Detectors through Sub-pixel Cross-coupling}
		
		\author{Sinuo Zhang\corref{a}}
		\cortext[a]{Equal contribution}
		\author{David-Leon Pohl\corref{a}}
		\author{Tomasz Hemperek\corref{b} and Jochen Dingfelder}
		%\cortext[c]{Corresponding author}
		
		\address{Physikalisches Institut, Rheinische Friedrich-Wilhelms-Universit{\"a}t Bonn, Nu{\ss}allee 12, 53115 Bonn, Germany}
		
		\begin{abstract}
		We present a concept to improve the spatial resolution of silicon pixel-detectors via the implementation of a sub-pixel cross-coupling, which introduces directional charge sharing between pixels. The charge-collection electrode is segmented into sub-pixels and each sub-pixel is coupled to the closest sub-pixel of the neighboring pixel. 
			
		Such coupling schema is evaluated for a model sensor design with $\SI{50}{\micro\meter}\times\SI{50}{\micro\meter}$ pixels and AC-coupled sub-pixels. A first-order SPICE simulation is used, to determine feasible coupling strengths and assess the influence on the charge-collection efficiency. The impact of the coupling strength on spatial resolution is studied with a dedicated simulation, taking into account charge-cloud evolution, energy-loss straggling, electronic noise, and the charge detection-threshold. Using simplifying assumptions, such as perpendicular tracks and no gaps between charge-collection electrodes, an improvement of the spatial resolution by up to approximately \SI{30}{\percent} is obtained in comparison to the standard planar pixel layout.
		\end{abstract}
		
	
	\end{frontmatter}
	
	%\linenumbers

%section 1	
	\section{Introduction}
	\noindent
		With high spatial resolution, fast timing, and good radia\-tion-hardness, pixel detectors are the preferred choice for the innermost particle trackers at various collider experiments~\cite{rossi2006pixel}. Thin planar silicon sensors are the state-of-the-art for the new generation of pixel detectors, featuring low material budget, low power consumption, and sufficient radiation tolerance. Multiple thin pixel-sensor designs are currently under development, such as \SI{100}{\micro\meter} and \SI{150}{\micro\meter} sensors for the hybrid pixel-detectors at the ATLAS and CMS experiments~\cite{Hamer:2665116, Contardo:2020886} and \SI{50}{\micro\meter} sensors for pixel detectors at the proposed Compact Linear Collider~\cite{Dannheim:2673779}.\\
		\indent
		Thinning of sensors while keeping the pixel pitch fixed, leads to a reduced number of detecting pixels per charge deposition (cluster size) and therefore to a deterioration of spatial resolution~\cite{AlipourTehrani:2133128}. The reason for this is decreased charge sharing caused by reduced lateral extension of the charge cloud. 
		After high levels of radiation, the charge-carrier trapping and the higher detection threshold due to elevated noise levels further limit the cluster size \cite{AlipourTehrani:2133128, Giurgiu:1122994}.
		%Especially, after high levels of radiation, small charge depositions due to charge-carrier trapping cannot be discriminated from noise, and clusters have mainly the size of one \cite{AlipourTehrani:2133128, Giurgiu:1122994}.
		It is possible to increase the cluster size by reducing the pixel pitch and thereby improving the spatial resolution \cite{Contardo:2020886}. However, the minimum pixel pitch is limited by space requirements for the integration of the charge digitization electronics. For standard hybrid pixel-detector designs the minimum bump-bond pitch imposes an additional restriction. A recent concept for a sensor design, named Enhanced LAteral Drift~(ELAD) sensors \cite{Velyka_2019}, provides a method to enhance the lateral electric field and consequently the charge sharing by using additional buried $\mathrm{n^+}$ and $\mathrm{p^+}$ implants. However, the production of such sensors requires a complex and dedicated process. 
		A different approach enhancing the spatial resolution is to make use of capacitive coupling between readout electrodes, an effect already extensively studied for strip detectors~\cite{kotz1985silicon, dabrowski1996study}.
		Recent work indicates the possibility of beneficial cross-coupling also for pixel detectors~\cite{WANG201810, Nachman_2019}.\\
		\indent
		In this paper, we propose a concept to enhance the spatial resolution of pixel detectors by introducing directional charge sharing between pixels through implementing sub-pixel cross-coupling. This is studied by introducing one possible sensor design and evaluating the spatial resolution with simulations. 
		The pixels in the design are segmented into directionally AC-coupled sub-pixels, where inter-pixel capacitances are used to enhance capacitive charge sharing. Herewith the spatial resolution is enhanced, while maintaining the number of readout channels.
		Such sensors can be produced in a CMOS process using high-resistive wafers and are therefore suitable for the high radiation environments of modern collider experiments~\cite{Pohl_2017}. 
		In this paper we study a sub-pixel geometry for $\SI{50}{\micro\meter}\times\SI{50}{\micro\meter}$ pixels, discuss the feasibility based on an implementation strategy, and evaluate the influence of different coupling strengths on the spatial resolution based on simulations.

%section 2
	\section{A model pixel-sensor design with sub-pixel cross-coupling}
	\label{DesignConcept}
\begin{figure}[h]
			\centering
			\includegraphics[width=0.8\linewidth]{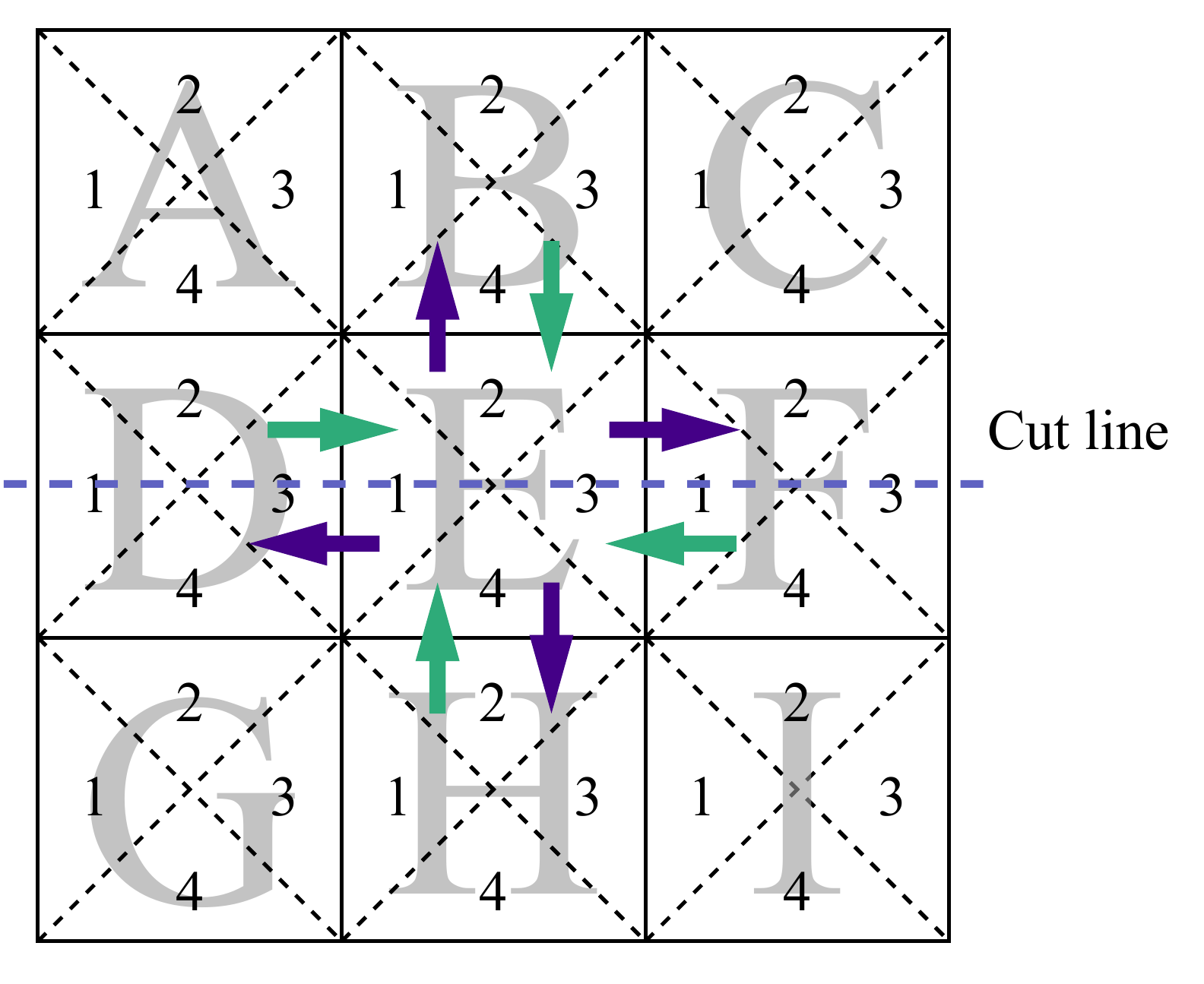}
			\caption{A 3x3 pixel matrix with sub-pixel segments in the form of triangles. Pixels are labelled with letters and the four sub-pixels with numbers. Arrows indicate the charge coupling between sub-pixels for the central pixel.}
			\label{pixmat}
		\end{figure}

\begin{figure}[h]
			\centering\includegraphics[width=1\linewidth]{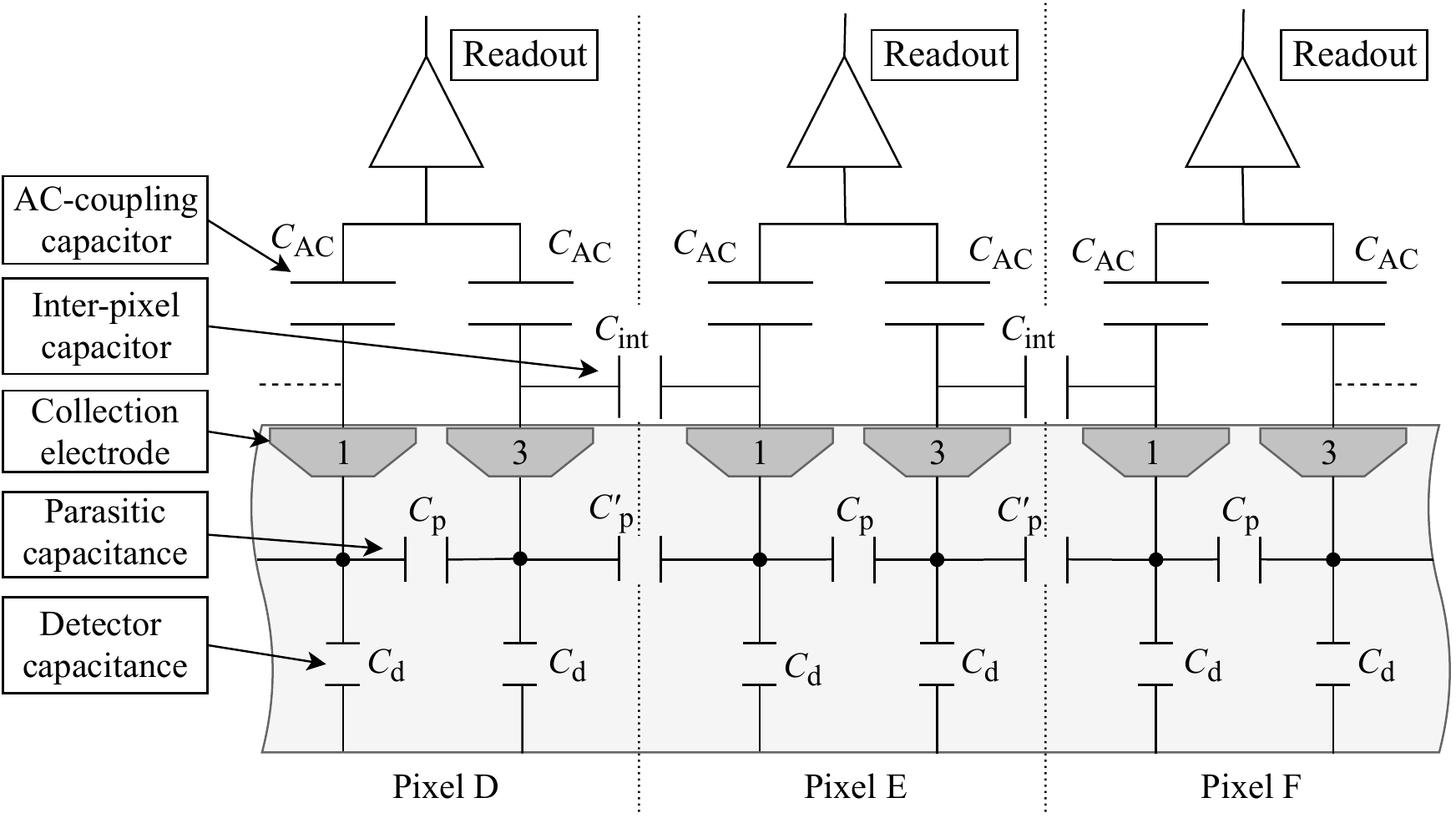}
			\caption{Equivalent circuit model for the sensor depicting inter-pixel and parasitic coupling capacitors in the sensor together with the pixel readout. The design is shown along the cut line given in Figure \ref{pixmat}.}
			\label{equicircuit}
		\end{figure}

	\noindent
		For the following investigations we consider a silicon sensor with square pixels, where the charge-collection electrode in this design is segmented into four sub-pixels in the form of triangles, as depicted in Figure~\ref{pixmat}.
		To study the influence of directional charge sharing on the spatial resolution, we employ one possible connection scheme, where each sub-pixel is AC-coupled to the closest sub-pixel of the adjacent pixel and all sub-pixels are AC-coupled to the readout of their corresponding pixel.	The equivalent circuit is illustrated in Figure \ref{equicircuit}, together with the readout electronics, for the pixels along the cut line depicted in Figure~\ref{pixmat}. For instance, sub-electrodes 1 and 3 (2 and 4 are not depicted) are AC-coupled to the readout channel via capacitors $C_\mathrm{{AC}}$, therefore the sum of their signals is received at the input of the readout electronics. The cross-coupling is realized via inter-pixel capacitances $C_\mathrm{{int}}$. Capacitive charge sharing also takes place between all sub-pixels via the parasitic capacitance $C_\mathrm{p}$ and $C'_\mathrm{p}$. The capacitance to the back plane $C_\mathrm{d}$ and parasitic capacitances can be confined to small values~($\sim\mathrm{fF}$) by controlling the size and depth profile of the electrode implants~\cite{Wang_2018}. To ensure good charge-collection performances, $C_\mathrm{{AC}}$ is in the order of $\SI{100}{\femto\farad}$ or even a few $\SI{}{\pico\farad}$. Therefore, the inter-pixel coupling capacitance must be sufficiently large to induce enough charge on the adjacent readout channel. Since $C_\mathrm{{int}}$, $C_\mathrm{d}$ and other parasitic capacitances are connected in parallel, the total inter-pixel capacitance is dominated by $C_\mathrm{{int}}$, which can cause an increase in the input capacitance.

%section 3	
	\section{Simulation}\label{sim}
	\subsection{Inter-pixel capacitance and charge sharing}
\begin{figure}[h]
			\centering
			\includegraphics[width=1\linewidth]{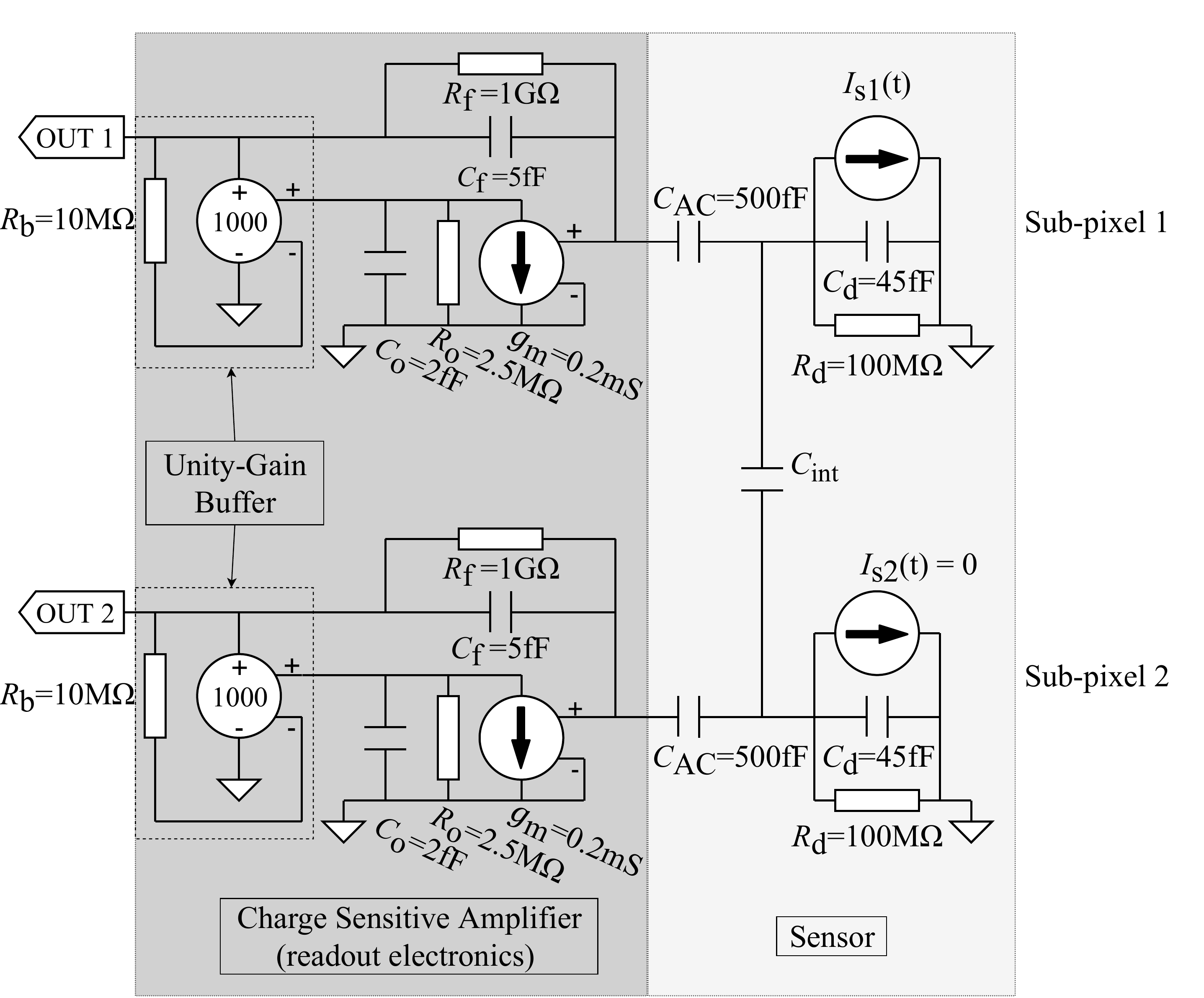}
			\caption{Simulation circuit of two sub-pixels which are coupled with an inter-pixel capacitor $C_\mathrm{int}$. The current pulse is injected in the sensor and integrated by the readout for a voltage signal.}
			\label{democircuit}
		\end{figure}

\begin{figure}[h]
			\centering
			\includegraphics[width=0.9\linewidth]{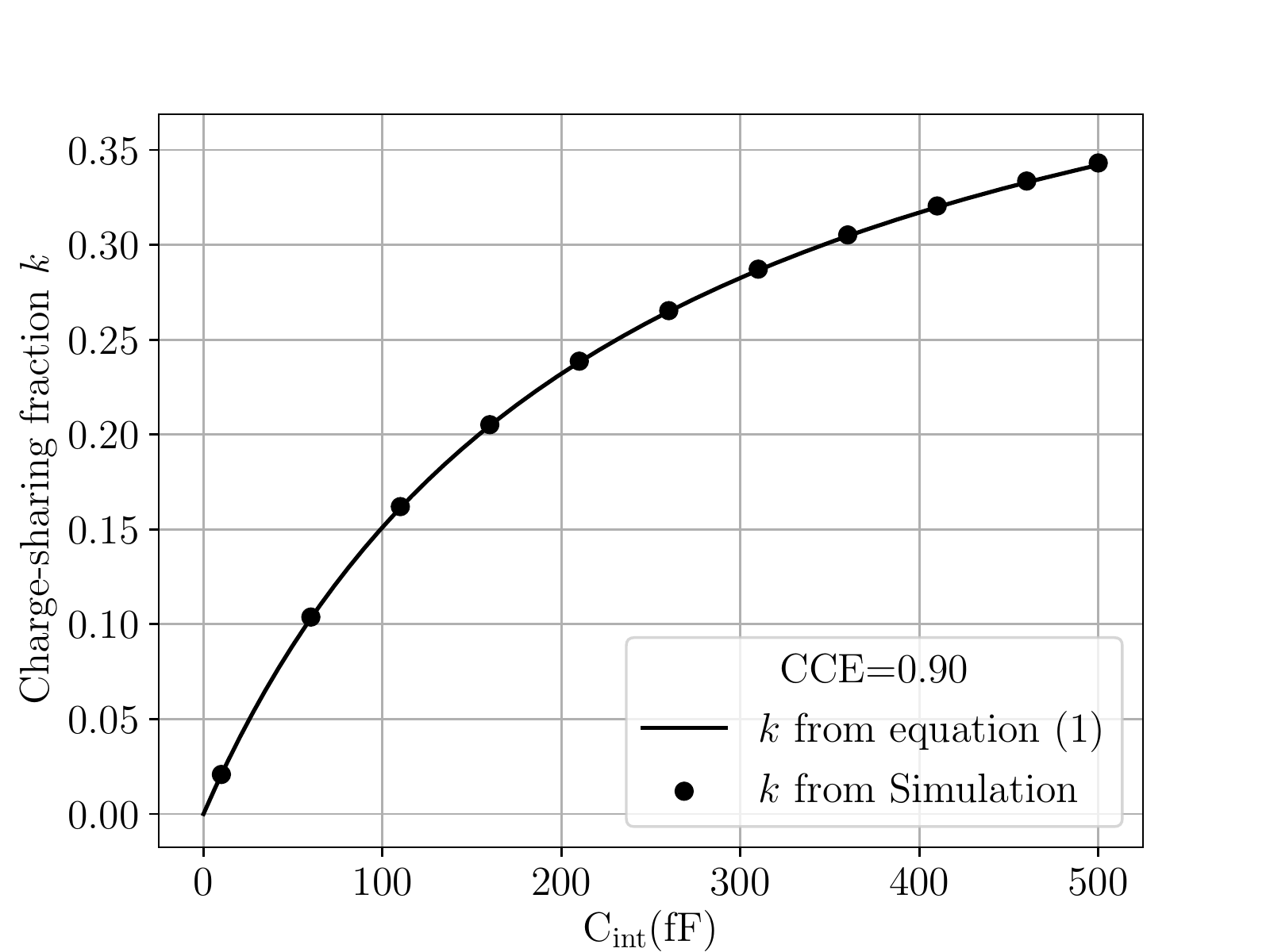}
			\caption{The charge-sharing fraction $k$ as a function of $C_\mathrm{int}$ is obtained for one sub-pixel current injection with a charge-collection efficiency of $90\%$. The analytical calculation from equation (\ref{QFcsa2}) is presented for comparison.}
			\label{csf}
		\end{figure}

		\noindent The capacitive charge sharing between coupled electrodes depends on the inter-pixel capacitance and is determined with transient simulations using SPICE~\cite{Nagel:M382}. As depicted in Figure \ref{democircuit}, a simplified, first-order model is implemented considering one pair of AC-coupled sub-pixels (e.g. sub-pixel 3 of pixel D and sub-pixel 1 of pixel E in Figure \ref{equicircuit}) and no parasitic capacitances.
		The sensor is represented by a parallel circuit containing the sensor capacitance $C_\mathrm{d}$ at full depletion, a resistor $R_\mathrm{d}$ for the DC-current working point, and a current source ($I_\mathrm{{s1}}$, $I_\mathrm{{s2}}$) that mimics the charge deposition of a particle via a current pulse. It has a rise time of $0.2~\mathrm{ns}$, peaks at \SI{0.7}{\micro\ampere} for \SI{1}{\nano\second} and has a fall time of \SI{1}{\nano\second}.
		The readout consists of a charge-sensitive amplifier~(CSA) with a feedback capacitor $C_\mathrm{f}$ and a large bleeder resistor $R_\mathrm{f}$. The CSAs are modelled using an ideal voltage dependent current source with transconductance $g_\mathrm{m}$, a capacitor $C_\mathrm{o}$ and a parallel resistor $R_\mathrm{o}$ leading to an open-loop gain of $a_\mathrm{o}=500$, hence an effective capacitance of the CSA as $C_\mathrm{{CSA}}=C_\mathrm{f}(a_\mathrm{o}+1) \approx \SI{2.5}{\pico\farad}$. Sensor and readout are AC-coupled using a capacitor $C_\mathrm{{AC}}=\SI{500}{\femto\farad}$, which together with the serial connected $C_\mathrm{{CSA}}$ delivers a total capacitance as $C_\mathrm{{CSA+AC}}\approx\SI{417}{\femto\farad}$.
		%With a realistic pixel input capacitance (CB ~ fF) about 90\% of the deposited charge is collected by by the readout.
		With the effective capacitance of the CSA~($C_\mathrm{{CSA}}$), the circuit in Figure \ref{democircuit} can be reduced to a capacitance net. Evaluation of parallel and serial capacitances leads to the following equation describing the charge $Q_\mathrm{2}$ in a sub-pixel for a charge $Q_\mathrm{0}$ deposited in its neighboring AC-coupled sub-pixel:
		\begin{linenomath}
		\begin{equation}
		Q_\mathrm{2}=Q_\mathrm{0}\underbrace{\dfrac{C_\mathrm{int}}{C_\mathrm{CSA+AC}+C_\mathrm{d}+2C_\mathrm{int}}}_{\mathrm{charge\text{-}sharing~fraction:}~k}\underbrace{\dfrac{C_\mathrm{CSA+AC}}{C_\mathrm{CSA+AC}+C_\mathrm{d}}}_{\mathrm{CCE}}~\mathrm{.}
		\label{QFcsa2}
		\end{equation}
		\end{linenomath}
		The charge-collection efficiency~(CCE) represents the portion of the \emph{total deposited charge} $Q_\mathrm{0}$ collected by all readout channels. It increases with the ratio of the effective capacitance of the CSA over the detector capacitance  and is approximately \SI{90}{\percent} for the given values.
		When considering the full capacitance network and not only the closest neighbour, the CCE depends also on $C_\mathrm{int}$ and decreases to approximately $75\%$ for $C_\mathrm{int}=\SI{150}{\femto\farad}$.
		The \emph{charge-sharing fraction} $k = Q_\mathrm{2}/(\mathrm{CCE}\cdot Q_\mathrm{0})$ indicates the portion of the collected charge that is shared to the coupled sub-pixel.
		%A DC-coupling readout is also applicable by using a CSA with smaller $C_\mathrm{CSA}$, and the charge sharing property can be acquired by substituting $C_\mathrm{CSA+AC}$ in (\ref{QFcsa2}) with $C_\mathrm{CSA}$.
		The coupling capacitance $C_\mathrm{int}$ is varied from $\SI{0}{\femto\farad}$ to $\SI{500}{\femto\farad}$, while injecting signal $I_\mathrm{{s1}}(t)$ into sub-pixel 1 and keeping $I_\mathrm{{s2}}(t)$ constant.
		The value of $k$ is determined by the output voltages of the CSAs as $V_\mathrm{OUT2}/(V_\mathrm{OUT1}+V_\mathrm{OUT2})$. As depicted in Figure \ref{csf}, a charge-sharing fraction $k$ up to approximately $0.35$ can be reached, which is consistent with the analytic equation~(\ref{QFcsa2}). Larger $k$ up to $0.5$ can be achieved through larger $C_\mathrm{int}$ and/or smaller $C_\mathrm{CSA+AC}$. The stated coupling capacitor values are reachable in CMOS processes (typically \SI{2}{\femto\farad\per\square\micro\meter}) given the assumed pixel pitch of \SI{50}{\micro\meter} in either dimensions.

	\subsection{Charge sharing and spatial resolution}
		\noindent To investigate the impact of charge sharing on spatial resolution, a dedicated simulation is created using a silicon sensor with $\SI{50}{\micro\meter} \times \SI{50}{\micro\meter}$ pixels, p-type bulk, and $\SI{100}{\micro\meter}$ thickness. Such design is consistent with the inner tracker upgrades for the ATLAS and CMS~\cite{Hamer:2665116, 7581861} experiments at the future High-Luminosity Large Hadron Collider. However, no gaps between the readout $n^+$-implantations are assumed, corresponding to a fill factor of 100\% and parallel electric field lines. 
		%Particles with perpendicular tracks through the sensor are simulated by modelling the charge deposition with considering the extension of the charge cloud.
		The particles are injected one by one at random positions across a $3 \times 3$ pixel matrix as illustrated in Figure \ref{pixmat}. For the sake of simplicity, we compare the case with inter-pixel coupling to the case without inter-pixel coupling disregarding the effect of delta-electrons in both cases, as we expect the performance of both types to suffer in a similar way~(\cite{Boronat:2002654} and citations therein).
		% cloud width ~\cite{KIMMEL2010334}
		The evolution of the charge cloud for a point-like charge deposition is considered using the continuity equation, following~\cite{Gatti:1987pi}. Since charge carrier densities from low-energy photons and minimum-ionizing particles are low, it is sufficient to consider diffusion only~\cite{BELAU1983253}. Hence, the evolution of the charge cloud is described by a normal distribution for the charge-density, whose time dependent width is given as
		\begin{linenomath}
		\begin{equation}
			\sigma(t) = \sqrt{2Dt} = \sqrt{\dfrac{2\,\mu\,k_\mathrm{B}\,T}{\si{\elementarycharge}}t}~\mathrm{,}
			\label{sigt}
		\end{equation}
		\end{linenomath}
		with temperature~$T$, electron mobility~$\mu$, and applying the ``Einstein-relation'' $D/\mu = k_\mathrm{B}T/e$. The drift time~$t$ for charge carriers deposited at depth $z$ in an over-depleted sensor ($V > V_\mathrm{dep}$) with thickness $d$
		% FIXME https://doi.org/10.1016/0167-5087(83)90591-4 citation
		is inserted into \eqref{sigt}, yielding~\cite{disserDLP}
		\begin{linenomath}
		\begin{equation}
			\sigma(z) = d \sqrt{\dfrac{k_\mathrm{B} T}{eV_\mathrm{dep}}}\sqrt{-\mathrm{ln}\left(1- \dfrac{2 V_\mathrm{dep} }{d(V+ V_\mathrm{dep})}z\right)}~\mathrm{.}
			\label{sigz}
		\end{equation}
		\end{linenomath}
		To describe the charge sharing from perpendicular tracks of minimum-ionizing particles, we use the projected charge cloud distribution onto the readout surface after drift. We assume that the projected charge cloud distribution from many charge depositions along a line-like charge deposition is also a two-dimensional Gaussian distribution with an effective width in $x$- and $y$-direction given as
		\begin{linenomath}
		\begin{equation}
		\bar{\sigma} = \dfrac{1}{d}\int_0^d \sigma(z) \mathrm{d}z \approx \SI{1.6}{\micro\meter}~\mathrm{.}
		\label{sigmabar}
		\end{equation}
		\end{linenomath}
		The calculated width follows from the parameters chosen in the simulation, assuming room temperature~($T=\SI{300}{\kelvin}$), a bias voltage of $V=\SI{80}{\volt}$, and a bulk resistivity of \SI{2}{\kilo\ohm\centi\meter} with a corresponding full-depletion voltage $V_\mathrm{dep}\approx\SI{50}{\volt}$. To estimate the charge collected per sub-pixel, the projected charge-cloud distribution after drift is integrated over the sub-pixel area:
		\begin{linenomath}
		\begin{align}
			&Q_\mathrm{sub\text{-}pix.} = \int_{\mathrm{area}} \mathrm{CCE}\cdot\rho\, \mathrm{d}x\,\mathrm{d}y\\
			\mathrm{with\ } &\rho = \dfrac{Q_\mathrm{0}}{2\pi\bar{\sigma}^2} \exp \left[ -\dfrac{(x-\mu)^2+(y-\nu)^2}{2\,\bar{\sigma}^2} \right]~\mathrm{.}
			\label{2dcloud}
		\end{align}
		\end{linenomath}
		Here, $Q_\mathrm{0}$ is the total charge deposit at position $(\mu, \nu)$, which varies between multiple charge depositions due to energy-loss straggling. The deposited charge is calculated as a random number from a Landau-like energy-distribution of minimum-ionizing particles~(\textit{MIP}) in a \SI{100}{\micro\meter} silicon layer as extracted from GEANT4~\cite{AGOSTINELLI2003250}.
		The capacitive charge sharing is simulated for each pair of coupled sub-pixels according to
		\begin{linenomath}
		\begin{equation}
			Q_\mathrm{tot,\, sub\text{-}pix.\, 1} = (1-k)\, Q_\mathrm{sub\text{-}pix.\, 1} +k\,\, Q_\mathrm{sub\text{-}pix.\, 2}
			\label{Cshare}
		\end{equation}
		\end{linenomath}
		and the total charge per pixel is calculated as the sum over the four sub-pixels.\\
		\indent
		The spatial resolution depends on the charge digitization parameters defined by the readout electronics, as e.g. shown in \cite{WANG201810}, and is therefore also included in simulation. We consider the most probable value of the Landau distribution $Q_\mathrm{0,MPV}$ as a reference for the total charge and assume a Gaussian noise with a standard deviation of $\SI{5}{\percent}\,Q_\mathrm{0,MPV}$ for each pixel. The detection thresholds are set to \SI{10}{\percent}\,$Q_\mathrm{0,MPV}$ and \SI{20}{\percent}\,$Q_\mathrm{0,MPV}$. Charge above threshold is detected as a hit and undergoes digitization with a charge discretization error of $\SI{10}{\percent}\,Q_\mathrm{0,MPV}$, which represents the minimum resolvable amount of charge. These values are comparable to the performance of existing pixel detectors, e.g., of the ATLAS and CMS experiments~\cite{Aad_2008,Chatrchyan:2009aa}.
		\begin{figure}[h]
			\centering
			\includegraphics[width=0.85\linewidth]{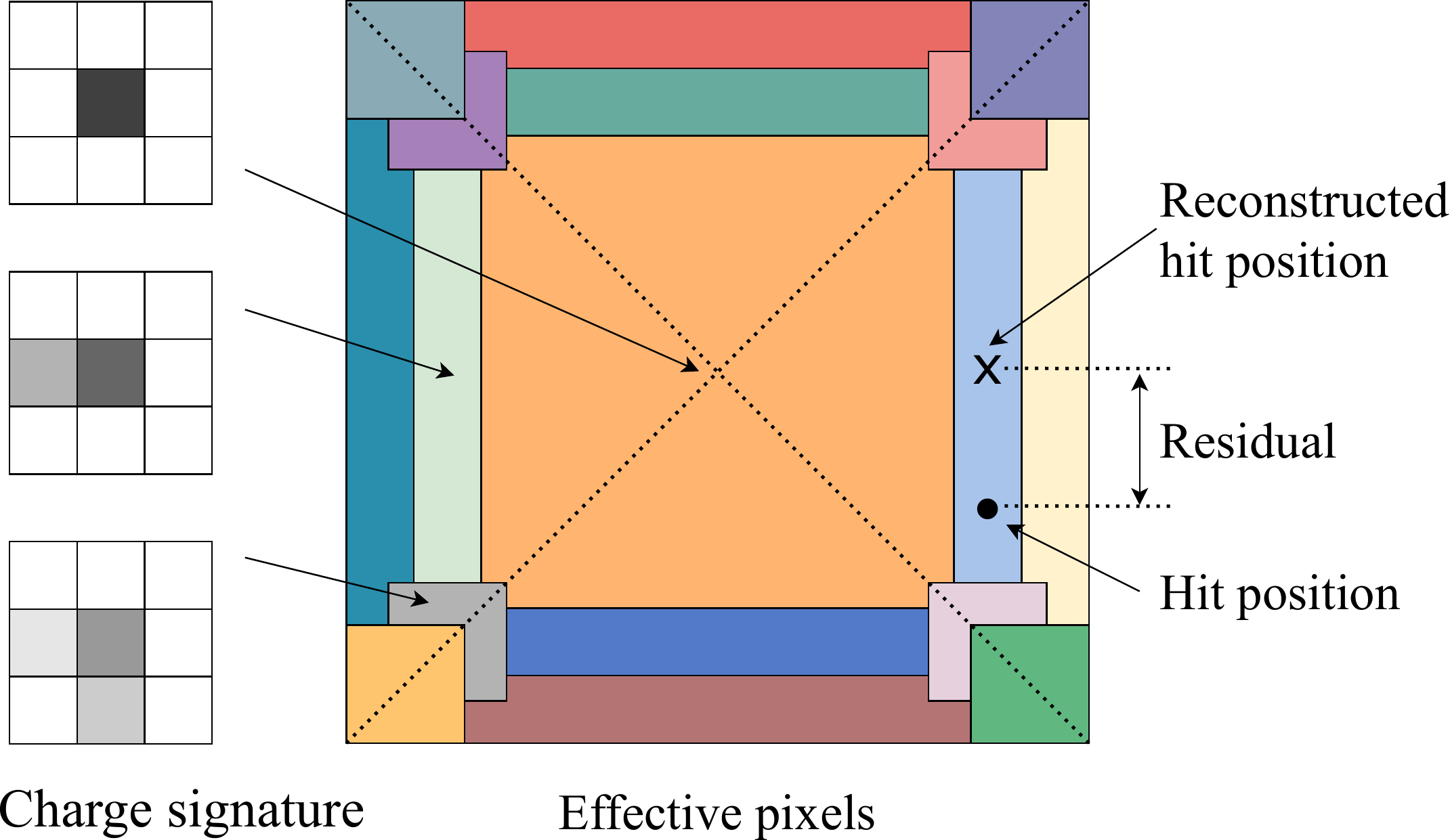}
			\caption{Illustration of the hit-reconstruction method through the example of a standard pixel sensor without cross-coupling. The right part shows the layout of effective pixels in one pixel. Effective pixels are shown as the colored areas and each represents a unique charge signature, which is schematically illustrated by the left part in grayscale.}
			\label{effpixDEMO}
		\end{figure}\\
		For hit-position reconstruction, we employ a template-based algorithm, similar to the technique introduced in~\cite{Swartz:1073691}. Cluster shapes and charge information of pixel hits are pre-computed
		as a function of the particle-hit position (referred to as \textit{templates} or \textit{charge signatures}) and then used during hit-position reconstruction.
		Pre-computation is done by simulating one million charge depositions from MIPs across the pixel matrix at random positions and calculating the corresponding charge of the nine pixels~(\textit{charge signature}).
		The charge signatures encode the hit positions in a certain area and depend on threshold, charge-sharing fraction, charge discretization error, and total charge deposit. From hit positions with the same charge signature an average position is calculated and assigned to the signature for reconstruction.
		Figure \ref{effpixDEMO} depicts different areas within the pixel with the same charge signature~(\textit{effective pixels}) that occur due to the finite charge resolution. For the depicted standard sensor pixel, without cross-coupled sub-pixels, most particle hits lead to a cluster size of one and are reconstructed at the center-of-gravity of the corresponding effective pixel~(\textit{central orange area}). Closer to the pixel boundaries, with more disjunct charge signatures, more effective pixels with smaller pitch occur, and the distance between reconstructed hit position and real hit position~(\textit{residual}) decreases. With the introduction of cross-coupled sub-pixels, more effective pixels with smaller pitch exist, decreasing the residuals and therefore increasing the spatial resolution. For the comparison between standard and sub-pixel sensor layout, the charge signatures for both designs are pre-computed for a varying total charge to consider energy-loss straggling.
		
		To assess the spatial resolution, one million charge depositions at different positions within a pixel are simulated. The charge signature after digitization is calculated and the position is reconstructed at the center-of-gravity of the corresponding effective pixel. Due to electronic noise the charge signature may not exist, in which case the reconstructed position is set to the pixel center.
		The RMS in x/y for all charge depositions $N$
		\begin{linenomath}
		\begin{equation}
			\mathrm{RMS} = \sqrt{\frac{1}{N} \sum_i^N (x_i - x_\mathrm{i, rec})^2}
			\label{resolution_simulation}
		\end{equation}
		\end{linenomath}
		is used to define the spatial resolution.
% section 4	
	\section{Results}
\begin{figure}[t]
			\center
			\includegraphics[width=1\linewidth]{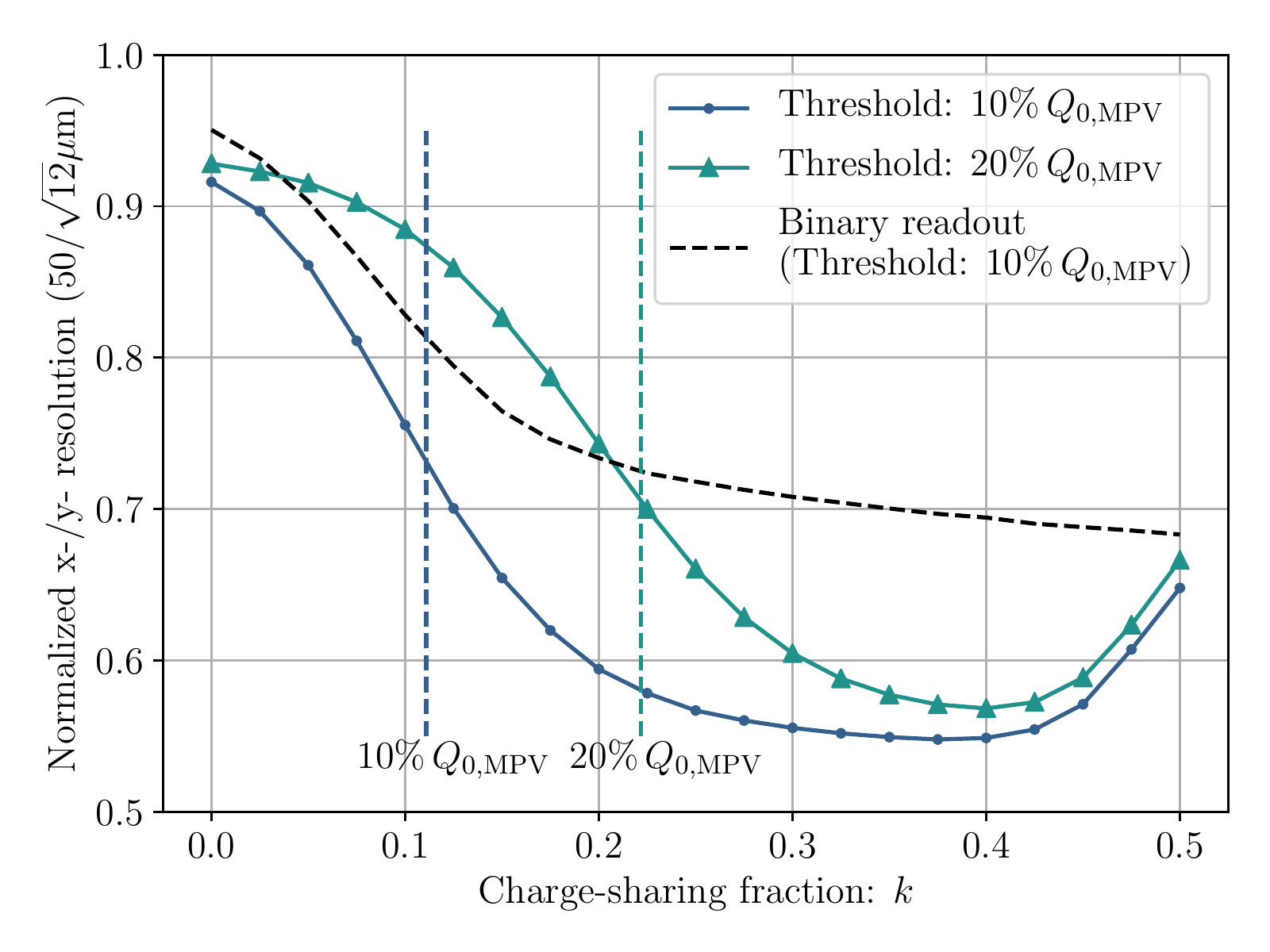}
			\caption{Normalized spatial resolution in x and y as functions of the charge-sharing fraction $k$. The results are normalized to $50/\sqrt{12}\,\mathrm{\mu m}$. The solid curves presents the cases with a threshold of $10\%\,Q_\mathrm{0,MPV}$ and $20\%\,Q_\mathrm{0,MPV}$. The dashed curve indicates the binary readout case with threshold of $10\%\,Q_\mathrm{0,MPV}$. The vertical lines indicate the cases ``$\mathrm{shared\,charge} = \mathrm{threshold}$''. }
			\label{resres}
		\end{figure}

\begin{figure}[t]
			\centering
			\includegraphics[width=1\linewidth]{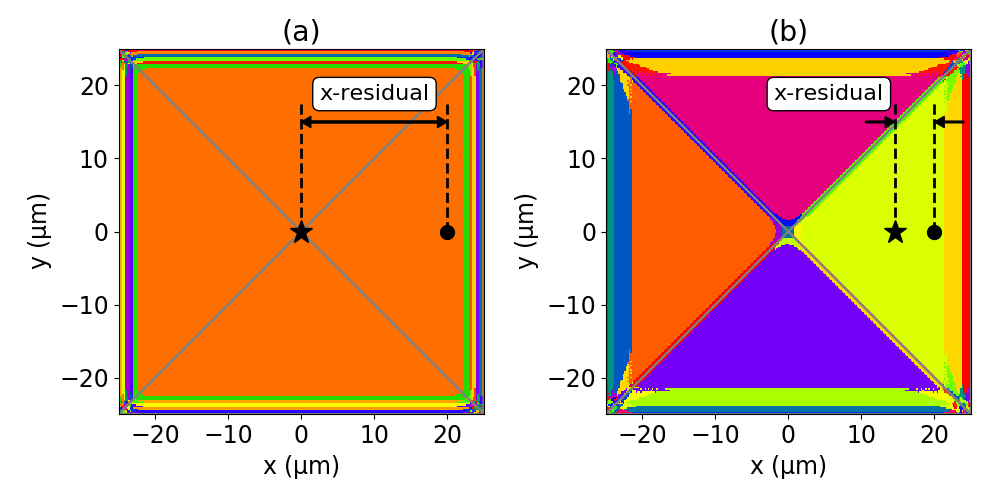}
			\caption{Layouts of effective pixels as different colored areas from a fixed total deposited charge for a charge-sharing fraction $k=0$~(a) and $k=0.25$~(b) with a threshold of $10\%\,Q_\mathrm{0,MPV}$. ``\ding{108}'' is the simulated particle's incident position, which is identical for both cases. ``\ding{72}'' represents the reconstructed position for the corresponding charge-sharing fraction. The residuals in x are illustrated as the distances between the dashed lines.}
			\label{effpixCOMP}
		\end{figure}

	\noindent
		The simulation results are presented in Figure~\ref{resres} for charge-sharing fractions $k$ up to $0.5$ and detection thresholds of $10\%\,Q_\mathrm{0,MPV}$ and $20\%\,Q_\mathrm{0,MPV}$ for the simple network at a CCE of 90\%. Without cross-coupling~($k=0$), charge is shared by diffusion. Given the small charge-cloud width in comparison to the pixel pitch, the cluster size is mostly one and a large effective pixel~(Figure~\ref{effpixCOMP} a) exists covering almost the total pixel area. Consequently, the spatial resolution is close to the maximal value bound by the pixel pitch~$p$:
		\begin{linenomath}
		\begin{equation*}
		\mathrm{RMS} = 
		\sqrt{\dfrac{1}{\mathrm{p}}\int_{-\mathrm{p / 2}}^{\mathrm{p / 2}} x^2\, \mathrm{d}x} %maybe not needed?
		= \dfrac{\mathrm{pixel\,pitch}}{\sqrt{12}} \sim \SI{14.4}{\micro\meter}~\mathrm{.}
		\end{equation*}
		\end{linenomath}
		Since this central effective pixel for cluster size one is independent of the charge resolution and dominates the $\mathrm{RMS}$ calculation, one does not observe large differences in spatial resolution between binary and non-binary readout.
		%FIXME: why important? would skip; This is more pronounced with the higher threshold.\\
		%\indent
		The spatial resolution improves with increasing charge-sharing fraction~(Figure \ref{resres}). A significant improvement over the standard design~($k=0$) by approximately $30\%$ occurs when the charge-sharing fraction exceeds the detection threshold, e.g. for threshold of $10\%\,Q_\mathrm{0,MPV}$ at $k\approx0.2$ and threshold of $20\%\,Q_\mathrm{0,MPV}$ at $k\approx0.3$. Here, charge signatures with larger cluster sizes encode the position of the cross-coupled sub-pixels leading to a finer division into effective pixels. The dominating four triangular effective pixels, as depicted in Figure~\ref{effpixCOMP}~(b), are the result of the sub-pixel electrode geometry.
		In the limiting case of binary readout these sub-pixels are only encoded in the cluster size which still gives an improvement of approximately $20\%$ at $k\approx0.2$, as shown in Figure \ref{resres}.
		When $k$ approaches $0.5$, the resolution degrades and approaches the resolution given by the binary readout, because the charge signature of the hits in adjacent pixels are the same. The alteration of charge, originating from noise and energy-loss straggling, can lead to the calculation of unknown or wrong charge signatures that in further consequence can lead to a worse position reconstruction. However, by choosing 1 million charge depositions to pre-compute the charge signatures the likelihood of having a signature during reconstruction that was not pre-computed is below \SI{2}{\percent}.
		
		
		\begin{figure}[t]
			\centering
			\includegraphics[width=1\linewidth]{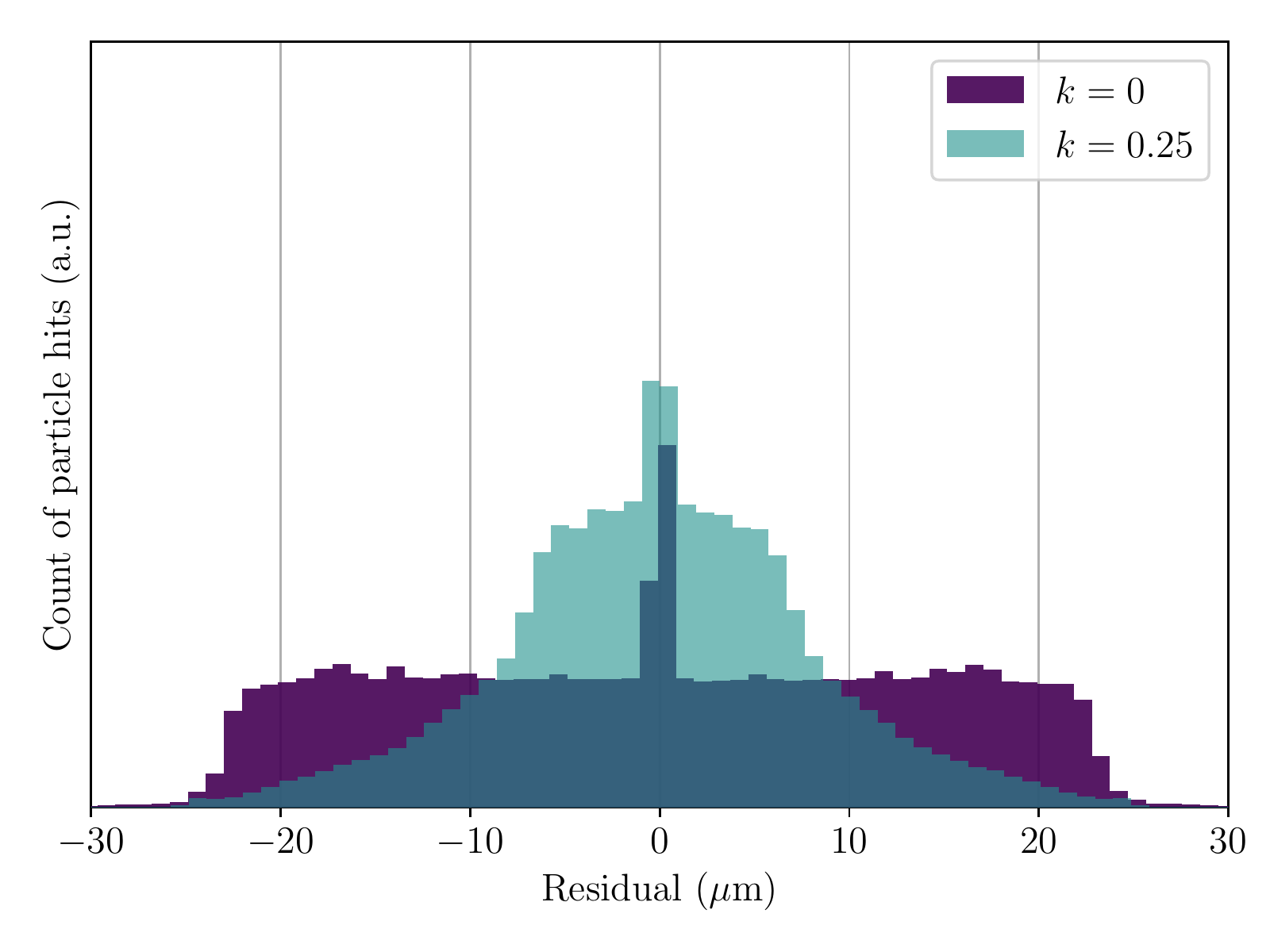}
			\caption{Residual distributions for a charge-sharing fraction $k=0$ and $k=0.25$ with a threshold of $10\%\,Q_\mathrm{0,MPV}$.}
			\label{xresid}
		\end{figure}

	\section{Conclusion}
	\noindent
	The concept of improving spatial resolution in pixel detectors by introducing directional charge sharing through sub-pixel cross-coupling is demonstrated using simulations. It offers a possibility to further enhance the spatial resolution while maintaining the pixel pitch and the density of readout electronics. The impact of charge-sharing on the spatial resolution is studied considering a $\SI{50}{\micro\meter} \times \SI{50}{\micro\meter}$ pixel sensor with \SI{100}{\micro\meter} thickness and assuming perpendicular particle tracks. An improvement in spatial resolution of approximately $30\%$ with respect to standard planar pixel sensor-layouts is obtained, independent of the detection threshold. The requirement is a sufficiently large charge-sharing fraction, so that the collected charge in adjacent sub-pixels is above the detection threshold. 
	
	A model coupling scheme using AC-coupled, triangular readout electrodes, is used as a possible implementation. Charge-sharing fractions up to $35\%$ are achieved with inter-pixel capacitors below \SI{500}{\femto\farad}, based on a first-order model. However, such coupling schema comes at the cost of a considerably increased input capacitances that, depending on the CSA design, worsens noise and timing performance or increases power consumption. This has a potentially negative impact on the in-time hit-detection performances.	
	There are alternatives for implementing the directional charge sharing, such as a different cross-coupling mechanism and sub-pixel geometry to be studied.
	Nevertheless, due to the interdependence of process and design in combination with the readout electronics, such performance figures are difficult to access with simulations, especially after radiation damage, requiring measurements on produced devices.

\section{Acknowledgement}
	\noindent
	We like to thank Konstantinos Moustakas, Tianyang Wang, and Hans Kr\"uger for their feedback concerning the SPICE simulation. This work was supported by the German Federal Ministry of Education and Research (BMBF) under grant number 05H15PDCA9.

%\section{Reference}
\bibliographystyle{elsarticle-num}
\bibliography{MAIN}
\end{document}